\documentclass[ reprint, nofootinbib,preprintnumbers,amsmath,amssymb]{revtex4}
\usepackage{varioref,exscale,latexsym,amsmath,amssymb}
\usepackage{graphicx}

\usepackage{graphicx}
\usepackage{slashed}
\usepackage{dcolumn}
\usepackage{bm}
\usepackage{hyperref}
\usepackage{color}
\usepackage{xcolor}

\newcommand{\beq}{\begin{equation}}
\newcommand{\eeq}{\end{equation}}
\newcommand{\bea}{\begin{eqnarray}}
\newcommand{\eea}{\end{eqnarray}}
\newcommand{\nn}{\nonumber}

\def\lsi{\raise0.3ex\hbox{$<$\kern-0.75em\raise-1.1ex\hbox{$\sim$}}}
\def\gsi{\raise0.3ex\hbox{$>$\kern-0.75em\raise-1.1ex\hbox{$\sim$}}}

\def\beq{\begin{equation}}

\def\eeq{\end{equation}}

\begin{document}
\preprint{ACFI-T20-03}

\title{{\bf Quantum causality and the arrows of time and thermodynamics}}

\medskip\

\medskip\

\author{John F. Donoghue${}^{1}$}
\email{donoghue@physics.umass.edu}
\author{Gabriel Menezes${}^{2}$}
\email{gabrielmenezes@ufrrj.br}
\affiliation{
${}^1$Department of Physics,
University of Massachusetts,
Amherst, MA  01003, USA\\
${}^2$Departamento de F\'{i}sica, Universidade Federal Rural do Rio de Janeiro, 23897-000, Serop\'{e}dica, RJ, Brazil}

\begin{abstract}
In the understanding of the fundamental interactions, the origin of an arrow of time is viewed as problematic. However, quantum field theory has an arrow of causality, which tells us which time direction is the past lightcone and which is the future. This direction is tied to the conventions used in the quantization procedures. The different possible causal directions have related physics - in this sense they are covariant under time-reversal. However, only one causal direction emerges for a given set of conventions. This causal arrow tells us the direction that scattering reactions proceed. The time direction of scattering in turn tells us the time direction for which entropy increases - the so-called arrow of thermodynamics. This connection is overlooked in most discussions of the arrow of time.
\\
\\
\\
Keywords: Causality, Arrow of time, thermodynamics, quantum field theory
\end{abstract}
\maketitle


\section{Quantum causality}

Eddington introduced the concept of the arrow of time - the one way flow of time as events develop and our perceptions evolve. He pointed out that the origin of such an arrow appears to be a mystery in that the underlying laws of physics (at least at the time of Eddington) are time symmetric and would work equally well if run in the reverse time direction. The laws of classical physics follow from the minimization of the action and are indeed time symmetric. This view has been beautifully captured by Carlo Rovelli \cite{Rovelli}, who writes: {\it ``The difference between past and future, between cause and effect, between memory and hope, between regret and intention...in the elementary laws that describe the mechanisms of the world, there is no such difference.''} But what picks out only those solutions running forward in time?

By now, there is a large literature on the arrow of time~ \cite{Rovelli, Halliwell:1994wq, Zeh, Carroll:2010zz, Coveny, Price:96, Hoerl:98, Dainton:13,Poidevin:19, Rovelli:2015aza}. Essentially all of the literature accepts the proposition that the fundamental laws of physics do not distinguish between past and future and could equally well be run backwards. There is also a recognition that the second law of thermodynamics {\em does} distinguish between these directions as it states that entropy cannot decrease in what we refer to as the future. This leads to the idea of a thermodynamic arrow of time. Many view this thermodynamic arrow as the origin of the passage of time, or at least of our consciousness of that passage.

Our point in this paper is that the basic premise of such reasoning is not valid in quantum theory. Quantum physics in its usual form has a definite {\em arrow of causality } - the time direction that causal quantum processes occur. Causality in quantum field theory is phrased in terms of commutators of field operators commuting at spacelike separations. This is used to tell us that events outside your past lightcone cannot influence your present physics. But in addition to the commutator condition, there is a directionality implied, which tells us what is the past lightcone and what is the future. This is encoded in the $i \epsilon$ in particle propagators. The usual choice implies that positive energy is propagated forward in time. If you change the sign of the  $i \epsilon$, you get the time reversed propagation. This is the arrow of causality.

These factors of $i$ can be traced back to factors of $i$ in the quantization procedure, such as using
$\exp[ +i S]$ vs $\exp[ -i S]$ in path integrals or using $+i\hbar$ vs $-i\hbar$ in canonical commutation rules. Complex conjugating the factors of $i$ there leads to the opposite factors of $ i \epsilon$ in propagators, as is appropriate for the time reversal operation. One obtains the same phenomena with either choice (neglecting small T violation), just with a different clock convention.

However with a choice of quantization rules, there is only one causal direction. Whichever choice you make applies to all reactions.

This contrasts with classical physics. If all that you are using are the equations of motion derived by minimizing the value of the action, you do not see that causal direction. Mixtures of retarded solutions and advanced solutions are also solutions of the equations of motion. There is no ingredient of the fundamental laws of classical physics which points to a direction of causality or time. That is the classical origin of the arrow of time discussion.

This arrow of causality is most visible in quantum field theory, where causality has long played a foundational role \cite{Fermi:32, GellMann:1954db, Eden:1966dnq, Maiani:1994zi, Coleman}. However, since usual quantum mechanics is a limiting case of quantum field theory, it must also have the same property. Part of this paper is devoted to showing how the same causal property appears in quantum mechanics. This can be concealed by conventions for how we count the flow of time, and associated conventions in quantum foundations. However  when the phrase ``fundamental laws of physics'' includes the rules for quantization, there is always only one time direction (for a given set of conventions) whose flow is compatible with quantum processes.

An example will illustrate our main point.
Consider a comparison of reactions of a style which are often mentioned as a test of time-reversal invariance, which we will imagine proceeding through a long-lived resonance,
\beq\label{timereversed}
A+B \to R\to C+D ~~~~~~~~vs ~~~~~~~C+D \to R \to A+B   \ \ .
\eeq
Here $A,~B,~C,~D$ are particles and $R$ is a resonance. These reactions are related by time reversal, as their matrix elements are complex conjugates of each other. It is often colloquially said that the second reaction is the backwards-in-time version of the first one. However, in reality both proceed forwards in time and the final state products emerge from the long-lived resonance only after a positive time. The equality of the transition probabilities is a consequence of the time reversal invariance of the underlying Lagrangian, but the fact that reactions run forward in time is a property of the arrow of causality of quantum physics. That this example invokes a long-lived resonance is not essential - it is done only to make the time ordering of the process manifestly evident. In reality all scattering reactions carry the same causal ordering.

Conventions play a role in all such discussions. In what might initially seem like a joke, we can appear to make these reactions ``run backwards in time'' by using a clock which counts down instead of up. We can make a clock which runs backwards. Or when using an hourglass the lower chamber reads time by an increasing amount of sand while the upper chamber reads time by a decreasing amount of sand. If we do this, then our resonance reactions will proceed in negative directions of this new time. This is less of a joke than it seems because the reversing of time is mathematically described by a substitution of a reversed time variable, i.e. $\tau = -t$, which is the substitution highlighted by the classical arrow of time puzzle. We need to be able to differentiate between having this substitution corresponding to just running our clocks backwards, and having physical processes running both directions in time.

The distinction highlighted by this ``joke" illustrates the difference between the arrow of causality and the arrow of time. There is a convention associated with the measurement of time, which is not a relevant distinction. However, what we really observe is that all elementary processes run in one direction only - that is causality. One can always choose to align one's clock in the direction of the causal action. But one does not find elementary processes running in both causal directions.  It is for this reason that it is better to refer to an arrow of causality rather than an arrow of time.

The causality relation applies to all reactions. For simple elastic scattering processes, for example
$\left(\pi\pi\right)_{\rm P-wave}\to \left(\pi\pi\right)_{\rm P-wave} $ there may be only a single initial state and single final state. However there are also many reactions with a single initial state and many possible final states. In these cases the scattering will lead to increased disorder. This tells us that the arrow of causality will also lead to to an arrow of thermodynamics. It is interesting that the macroscopic and classical arrow of time follows from the underlying causal quantum processes. To the best of our knowledge the disentangling of real causal physics from conventions has not been presented before in connection with the arrow of time. The discussion here is an elaboration of the comments which we presented in Ref. \cite{Donoghue:2019ecz}.

\section{The Laws of Physics}

When we talk about the fundamental ``laws of physics'', what do we mean? For classical physics, it is best to describe these by the Principle of Least Action and the use of Lagrangians defining the action. This is because Noether has taught us that invariances of the action reveal symmetries of the world. The Lagrangians of the fundamental interactions are (mostly) invariant under time reversal \footnote{There can be small violations of time reversal symmetry in the Standard Model Lagrangian, but this has nothing to do with the arrow of time.}. Simply from this fact alone, we know that solutions to the resulting classical equations of motion will then be equally possible in both time directions.

In order to succinctly state the classical puzzle about the arrow of time, let us consider the Principle of Least Action. The action is defined via the Lagrangian
\beq
\hspace{-2.4mm}
S= \int dt ~L\left(x,\frac{dx}{dt}\right) = \int dt \left[ \frac12 m \left(\frac{dx}{dt}\right)^2 - V(x)\right]
\eeq
where $m$ is the mass of the particle and $V(x)$ is the potential. We are interested in the direction of time, so let us consider two choices. We will use $t$ for the usual time which runs in an increasing direction, and use $\tau =-t$ for a time coordinate which runs in the reverse direction. The action is invariant under this change
\beq
S= \int dt ~L\left(x,\frac{dx}{d\tau}\right) = \int d\tau \left[ \frac12 m \left(\frac{dx}{d\tau}\right)^2 - V(x)\right]  \ \ .
\eeq
Likewise the equations of motion which follow from this Lagrangian are the same for either $t$ or $\tau$,
\beq
\frac{d}{dt} \frac{\partial L}{\partial v}- \frac{\partial L}{\partial x}= \frac{d}{d\tau} \frac{\partial L}{\partial \tilde{v}}- \frac{\partial L}{\partial x} = 0
\eeq
where $v = dx/dt$ and $\tilde{v} = dx/d\tau$. This is the origin of the classical puzzle about the arrow of time. For every solution using $t$, there is an identical solution using $\tau$. {Indeed, if $f(t)$ is a solution to the equations of motion, then so is $f_+(t)= (f(t)+f(-t))/2$. A similar analysis can be done for classical fields. }Classical physics does not contain any information distinguishing the direction of time.

{ However, quantum physics is different. Here, }the ``laws of physics'' not only include the Lagrangians and their subsequent equations of motion, but also the quantization procedures. The quantization rules, no matter how they are phrased, involve factors of $i$. These factors have an additional role in the discussion of time reversal. We can see this simply in the canonical quantization rules. The canonical momentum formed with $\tau$
\beq
p_\tau = \frac{\partial  L}{\partial (\partial_\tau x)}
\eeq
is the negative of the canonical momentum formed with $t$. This implies that the quantization conditions
\beq
[x,p_t] = i\hbar ~~~~{\rm vs} ~~~~[x,p_\tau] = i\hbar
\eeq
also are different by a minus sign. Likewise in path integral quantizationn one can use $\exp[ +iS]$ or
$\exp[-iS]$. What is the implication of these differences? We will show that the arrow of causality emerges from the factors of $i$. The time reversal operation in quantum physics is anti-unitary, that is it involves complex conjugation. These factors of $i$ then change under time reversal, so that the classical notion is modified. It is still the case that the time reversed situation is equivalent to the initial one, but both cannot be simultaneously valid. Probably the focus on the equations of motion alone and not on the quantization procedures is what has led astray many of the discussions of the arrow of time.

\section{Causality in quantum field theory}\label{QFT}

We start with the path integral treatment of quantum field theory, because this is the most direct illustration of causality. We will subsequently turn to canonical quantization and a discussion of conventions in order to extract the same physics.

As we proceed, it is important to separate what is convention from what is the intrinsic physics.
The bedrock convention needed here is that kinetic energy is positive. There is no point to entertaining any other convention. The rest mass and total energy of particles is then also positive. In our sample reactions of Eq.~(\ref{timereversed}) both initial states come with positive energy and the resonance $R$ has an energy which is sum of the initial energies.
We will be interested in following what happens to positive energy states.

Causality has been a major theme of quantum field theory from its early days \cite{Fermi:32, GellMann:1954db, Eden:1966dnq, Maiani:1994zi, Coleman}. The causal structure was used to formulate dispersion relations, and the analytic properties of amplitudes were heavily explored. One of the conditions for causal behavior is that commutators of field operators must vanish for spacelike separations,
\beq\label{causality}
[{\cal O}(x), {\cal O}(x')] = 0 ~~~{\rm for} ~~(x-x')^2 <0.
\eeq
such that causal influences are restricted to the past lightcone. However, this also requires the differentiation of past and future. The lesson of earlier work is that the direction of causality is encoded in a subtle feature of the particle propagation, namely the factors of $i\epsilon$ (with $\epsilon$ being an infinitesimal positive quantity) which determine the analytic structure of amplitudes.

\begin{figure}[htb]
\begin{center}
\includegraphics[height=60mm,width=80mm]{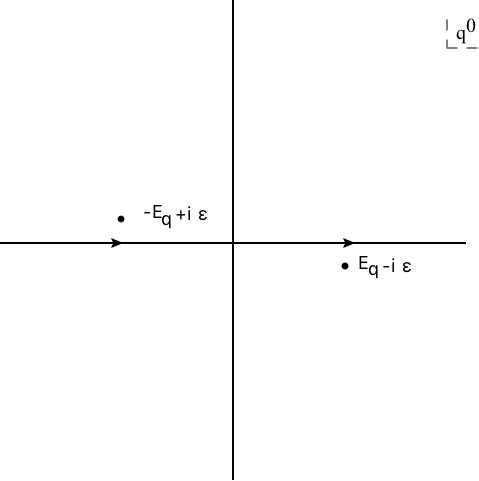}
\caption{The poles with the usual factors of $i\epsilon$ relevant for the evaluation of the Feynman propagator. When using the alternate propagator $D_{-F}$ these poles appear reflected on the other side of the real axis.}
\label{feynman}
\end{center}
\end{figure}

\begin{figure}[htb]
\begin{center}
\includegraphics[height=28mm,width=85mm]{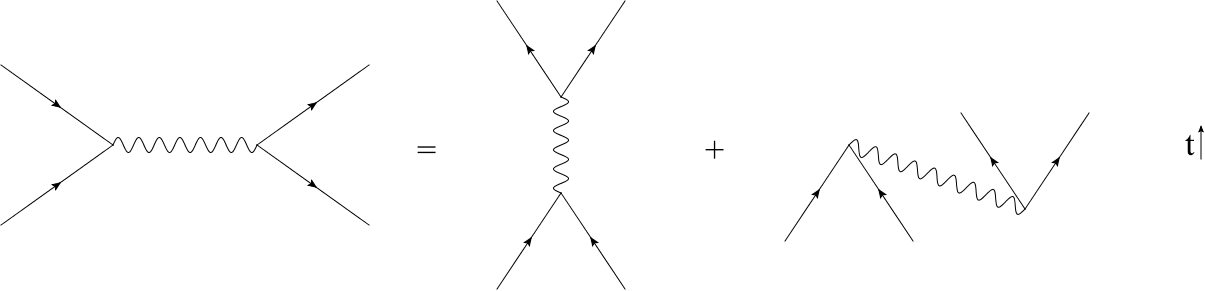}
\caption{The simple Feynman diagram on the left is decomposed into two time ordered diagrams.  With standard conventions, positive energy intermediate states are propagated forward in time. }
\label{timeorderings}
\end{center}
\end{figure}

The Feynman propagator governs the exchanges of particles in scattering amplitudes. With standard conventions, in momentum space it has the form (with $\hbar = c = 1$)
\beq
iD_F(q) = \frac{i}{q^2-m^2+i \epsilon}  \ \
\eeq
where our metric for four-vectors is $(+,-,-,-)$. The $i\epsilon$ is important. When we Fourier transform the propagator to coordinate space we can do the $q_0$ integration using contour integration with the poles indicated in Fig.~\ref{feynman}. The result includes both forward and backward propagation in time as in Figure \ref{timeorderings}, with
\beq
iD_{F}(x) = D_{F}^{\textrm{for}}(x) \theta (t) + D_{F}^{\textrm{back}}(x) \theta (-t)
\eeq
with
\beq
D_{F}^{\textrm{for}}(x) = \int \frac{ d^3q}{(2\pi)^3 2E_q} e^{-i(E_qt - {\bf q} \cdot {\bf x})}
\eeq
where $E_q = \sqrt{{\bf q}^2 + m^2}$ and $D_{F}^{\textrm{back}}(x) = (D_{F}^{\textrm{for}}(x))^*$. For our purposes here the most important aspect is that positive energy is, i.e. $e^{-iE_q t}$, is propagated forward in time. The scattering of positive-energy particles then carries this directionality.

If the propagator describes a normal unstable resonance, as in our example of Sec. 1, there is a finite imaginary part, $i\epsilon\to i\gamma$ with $\gamma >0$, such that
\beq\label{resonance}
iD_F(q) = \frac{i}{q^2-m^2+i \gamma}  = \frac{i}{q^2 - (M -i \Gamma/2)^2}\ \
\eeq
where $\gamma = M \Gamma$. The propagation forward in time exhibits exponential decay
\beq
D_{F}^{\textrm{for}}(x) = \int \frac{ d^3q}{(2\pi)^3 2E_q} e^{-i(E_qt - \vec{q}\cdot\vec{x})}e^{-\gamma t/2E_q}
\eeq
Again it is positive energy which is propagated forward in time.

The propagator demonstrated the existence of an arrow of causality. If we look at the reactions given in Eq.~ (\ref{timereversed}), both initial states carry positive energy. The scattering reaction then has a causal direction for the flow of that energy. When there is a long lived resonance involved, the resonance excitation, as in Fig.~\ref{timeorderings}, propagates the positive energy forward in time until it decays.  This is the physical origin of the time asymmetry which we described in the introduction.

However, we need to understand where this propagator came from, and why there is this directionality. In a path integral treatment we quantize using $\exp[{iS}]$ where $S$ is the action.  For a scalar field in the presence of an external source $J(x)$ we use
\bea\label{usual}
Z_+[J] &=& \int [d\phi ] e^{+ i S(\phi, J)}  ~~~~~~~~\fbox{1}
\nn\\
&=& \int [d\phi ] e^{+ i \int d^4 x [\frac12 (\partial_\mu \phi \partial^\mu \phi - m^2 \phi^2 )+ J\phi ]}  \ \ .
\eea
It will be seen that this is the origin of the usual propagator. However, we could also consider the case with the opposite factor of $i$
\beq
Z_-[J] = \int [d\phi ] e^{- i S(\phi, J)}  ~~~~~~~~\fbox{2} \ \ .
\eeq
What happens when we compare these two choices?

In these path integrals, we are confronted with Gaussian integrals, but with an imaginary argument. In order to make this well-defined we add a term
$ \pm  i \epsilon\phi^2/2$ to the Lagrangian density. When combined with the overall factor of $\pm i$, this provides a damping factor $\sim e^{-\epsilon \int d^4 x \phi^2/2}$ in the path integral. The two path integrals can be solved by defining
\begin{equation}
\begin{aligned}
\phi^{\prime}(x) &=\phi(x)+\int d^{4} y D_{\pm F}(x-y) J(y) \\
i D_{\pm F}(x-y) &=\int \frac{d^{4} k}{(2 \pi)^{4}} e^{-i k \cdot(x-y)} \frac{\pm i}{k^{2}-m^{2}\pm i \epsilon} \\
\left(\square_{x}+m^{2}\right) D_{\pm F}(x-y) &=-\delta^{(4)}(x-y)  \ \ .
\end{aligned}
\end{equation}
The path integral over all $\phi'$ is the same as that over $\phi$,
\beq
\int [d\phi'] =\int [d\phi]   \ \ .
\eeq
Using this one obtains
\beq
\hspace{-0.5mm}
Z_\pm [J] = Z[0] \exp\left\{-\frac12\int d^4x d^4 y J(x) ~iD_{\pm F}(x-y) J(y) \right\}
\eeq
with
\beq\label{plusminus}
i D_{\pm F}(x-y) = \int \frac{ d^4q}{(2\pi)^4} e^{-iq\cdot(x-y)} \frac{\pm i}{q^2 - m^2 \pm i \epsilon}   \ \ .
\eeq
The propagator with the plus sign $D_{+F}$ is just the usual Feynman propagator. The other propagator $D_{-F}$ is similar but with different analyticity properties, having poles in the complex plane being across the real axis from the usual case. Using these poles, it can be decomposed into time-ordered components. In this case the positive energy modes flow backwards in time
\beq
i D_{-F}(x) = D_{-F}^{\textrm{for}}(x) \theta (t) + D_{-F}^{\textrm{back}}(x) \theta (-t)
\eeq
with
\beq
D_{-F}^{\textrm{back}}(x) = \int \frac{ d^3q}{(2\pi)^3 2E_q} e^{-i(E_qt - \vec{q}\cdot\vec{x})}
\eeq
and negative energy forward in time $D_{-F}^{\textrm{for}}(x) = (D_{-F}^{\textrm{back}}(x))^*$. This is the time-reversed version of the usual propagator.

The propagator is used to generate the Feynman rules of the interacting theory. These rules describe the Feynman diagrams for scattering and decay amplitudes. If one uses $D_{-F}$ instead of the usual propagators, the resulting amplitudes are time reversed. The usual properties of field theory apply, but with a different arrow of causality. Both options for propagation are causal and have a direction of causality - they differ only on the convention chosen for the measurement of time.

{We note that the choice of $\pm i\epsilon$ was not driven by a desire to impose a preferred direction of time or causality. The direction was an output rather than an input. }

We also see that this difference is in accordance with the anti-unitary character of time-reversal symmetry. While the Lagrangian may be time-reversal invariant, the full path integral changes because we form it using $e^{iS}$. Under time-reversal, $Z_+$ is turned into $Z_-$, reversing the direction of the arrow of causality. Both versions of the quantum theory are equally causal, but with different directions of causality. The phenomena which emerge will be the same, but with a different convention of clock direction. However, the key lesson is independent of conventions. There is an intrinsic asymmetry in the elementary quantum reactions for scattering and decay of positive energy particles. These happen in one preferred direction, but not in the other. They carry an arrow of causality. The direction is ultimately tied to a factor of $i$ in the quantization condition.

The different directionality is also present in canonical quantization of fields. There the canonical momentum is derived from the Lagrangian
\beq
{\pi}= \frac{\partial {\cal L}}{\partial (\partial_t\phi)}
\eeq
and the commutation rule
\beq\label{usualcomm}
[\phi(t,x), { \pi}(t, x')] = i \hbar \delta^3(x-x')
\eeq
is imposed. This leads to the usual theory, and so is equivalent to case $\fbox{1}$, Eq.~(\ref{usual}).
However there could be a second choice with the factor of $i$ reversed
\beq\label{unusualcomm}
[\phi(t,x), { \pi}(t, x')] = -i \hbar \delta^3(x-x')  \ \ .
\eeq
This can be seen to be the time-reversed version of the original theory in the following way. We could have used the time variable
$\tau =-t$ instead, and found the corresponding canonical momentum
\beq
\bar{\pi}(\tau, x)= \frac{\partial {\cal L}}{\partial (\partial_\tau \phi)}  \ \ .
\eeq
It obvious that $\bar{\pi}(\tau, x) = -\pi(-t,x)$, so that the unusual commutation rule of Eq. \ref{unusualcomm} is just equivalent
to what would have the normal one for $\bar{\pi}$,
\bea\label{unusualcommtau}
&[\phi(\tau,x), {\bar \pi}(\tau, x')] = i \hbar \delta^3(x-x') \nonumber \\
& \Rightarrow
[\phi(-t,x), { \pi}(-t, x')] = -i \hbar \delta^3(x-x')  \ \ .
\eea
Despite the unconventional signs, the unusual commutation rule is the same theory, just with a change of the clock direction, as was choice $\fbox{2}$ above. Further development of these two options will lead to causal quantum theory with differing clock directions.

\section{Living in a causal world}\label{covariance}

How would we know if we were living in a world where quantum propagation was derived using $\exp[-iS]$ instead of $\exp[+iS]$, or with the alternate canonical quantization rules? We have seen that the unconventional alternate corresponds to propagation backwards in time compared to the conventional choice. Does this mean that we would see resonances decaying before they are produced?

It is here that the distinction between causal direction and time direction becomes significant. In a causal world, reactions flow in a single time direction, but not in the reverse direction. However, it is our convention whether we label that time direction that of increasing time or of decreasing time, by choosing how we measure the passage of time on our clocks. It does seem natural to choose the time direction to increase in the direction of causal action, but that is a choice.

Both the conventional and unconventional choices then lead to the same physical situation. Elementary reactions all occur in one time direction only, and we can use the clock of our choice. In this sense, the laws of physics are {\em time-covariant} but not {\em time-symmetric}. The operation of time reversal changes the factors of $i$ in the quantization rules, so that these rules are not invariant. But equivalent physics would be found with either set of rules with either choice of convention.

\section{Tests of Time reversal Symmetry}

In practice there is a small violation of time reversal symmetry in our fundamental theories. This is not the origin of the arrow of time. Nevertheless, the analysis of time reversal violation is interesting due to the interplay of causality and time symmetry.
Time reversal violation occurs due to complex phases (i.e. factors of $i$) which appear in the fundamental Lagrangian. These change sign under the anti-unitary time reversal operation. Not all such phases produce genuine T-violation, because some only produce overall phases in amplitudes or equivalently can be absorbed into the definitions of physical states. However, there can be residual factors of $i$ which are genuinely odd under the time reversal operation.

There are two points associated with T-invariance and causality which are not often stated. The first is that because we cannot physically run the reactions backwards in time, we need to compare two related reactions running forward in time. We are trying to isolate the effects of the T-violating phases from the Lagrangian, rather than actually reversing the flow of time. This complicates the analysis of T violation. The second point is that, since all measurements are real numbers, the T-violating factors of $i$ need to interfere with other factors of $i$ in order generate results which are real. Because T-violation is weak, in practice all of our measures of it are linear in the T-violating phases, so that this interference needs to occur with normal factors of $i$ in the amplitudes. These in turn can be ultimately traced back to the factors of $i$ in the quantization conditions because that is the place where the $i$ enters physics.

Consider first the spin asymmetry in neutron beta decay,
\beq
D\sim \mathbf{S}_n \cdot \mathbf{p}_e\times \mathbf{p}_\nu  \ \
\eeq
where $S_n$ is the initial spin of the neutron and the momenta are those of the final-state products. Because the momenta and spin are odd under time reversal, this asymmetry appears to be also. One would naively expect that comparisons of decays with $D$ positive and $D$ negative could serve as tests of T-invariance. However, it is well known that this is not the case. The signal can also be generated by final-state interaction (fsi) phases. What is wrong with the naive reasoning?  It is because one is not actually reversing time. The fsi phases can be traced back to the factors of $i\epsilon$ in propagators\footnote{To the best of our knowledge, the first to use this phrasing was E. Derman in Ref. \cite{Derman:1976rx}}, which we have stressed is related to the causal direction of reactions. The neutron decay for both $D$ positive and $D$ negative runs forward in time. In this situation, interactions in the final state can generate non-zero imaginary parts in the decay amplitude. It is the non-zero final state interaction phases which generates a non-zero expectation value for $D$ even in a T invariant theory. These imaginary parts can be calculated using the Cutkosky cutting rules \cite{Cutkosky:1960sp}, which in turn follow from the factors of $i\epsilon$ in the propagators\footnote{In the case of neutron decay the final state interactions are electromagnetic, involving the final proton and electron, and so can be calculated with reasonable accuracy. A signal at a level larger than the calculated fsi effect could still be attributed to T violation. In the Standard Model the expected T violation is much smaller than the fsi effect.}.

This dependence on final state interaction phases can also be seen in tests of direct CP violation, which by the CPT theorem are related to T violation. Consider for example the decay $K^0\to\pi^+\pi^-$, which can occur through either an $I=0$ or $I=2$ final state. The final state rescattering in these channels are described by the phase shifts $\delta_{0,2}$, and there can be weak T-violating phases which we can call $\phi_{0,2}$. The overall amplitude is then
\beq
A(K^0\to\pi^+\pi^-) = A_0 e^{i\delta_0} e^{i\phi_0} + A_2 e^{i\delta_2} e^{i\phi_2}
\eeq
where $A_{0,2}$ can be taken as real numbers. The corresponding decay of a $\bar{K}^0$ is then
\beq
A(\bar{K}^0\to\pi^+\pi^-) = A_0 e^{i\delta_0} e^{-i\phi_0} + A_2 e^{i\delta_2} e^{-i\phi_2} \ \ .
\eeq
Note that the T-odd phases change sign but that the final state phase shifts $\delta_i$ remain the {\em same} sign. This follows from the final state evolving in the same time direction with the same factors of $i\epsilon$, while the weak phases change because the Hermetian conjugate of the interaction Lagrangian is involved. In practice, the analysis of kaon CP violation is complicated by kaon mixing, but if that were not the case we would have a rate difference proportional to
\bea
R(K^0\to\pi^+\pi^-) - R(\bar{K}^0\to\pi^+\pi^-) &\sim& A_0A_2 \sin(\delta_0-\delta_2)
\nn\\
& \times& \sin (\phi_0-\phi_2)  \ \ .
\eea
This illustrates that in this process, the imaginary parts of the T-odd Lagrangian interfere with the phases formed in the propagation of the final state.

As a further example, let us consider what is often cited as the classic demonstration of T-violation, the difference between the oscillation of $K^0 \to \bar{K}^0$ versus the reverse $\bar{K}^0 \to {K}^0$ \cite{Kabir:1970ts, AlvarezGaume:1998yr, Adler:1995xy}. The observable is the rate asymmetry
\beq
A(t) = \frac{R\left(K^0(t=0) \to \bar{K}^0(t)\right)- R\left(\bar{K}^0(t=0) \to {K}^0(t)\right)}{R \left(K^0(t=0) \to \bar{K}^0(t)\right)+ R\left(\bar{K}^0(t=0) \to {K}^0(t)\right)}
\eeq
as a function of time. This is a variation on the propagation of a long-lived resonance, here treated as a coupled two-channel problem with the channels being the flavor eigenstates $K^0$ and $\bar{K}^0$. In the CPLEAR experiment \cite{Adler:1995xy} the initial resonant state is produced at $t=0$ by
\beq
p\bar{p}\to K^-\pi^+K^0  ~~~~{\rm or }~~~~p\bar{p}\to K^+\pi^-\bar{K}^0 \ \ .
\eeq
The final flavor eigenstate is tagged at time $t$ by the weak decay
\beq
K^0 \to \pi^-  e^+  \nu ~~~~{\rm vs} ~~~~\bar{K}^0 \to \pi^+   e^- \bar{\nu}  \ \ .
\eeq
While one oscillation is in some sense the other oscillation run backwards, in practice they both run forward in time. That is a reflection of the causal property of the quantum theory. However, the T-violating phases from the Lagrangian will in general give a phase difference to the two reactions, with one carrying some phase $e^{+i\phi(t)}$ and the other the phase $e^{-i\phi(t)}$. These phases interfere with the $e^{-iEt}$ factor that arises because both reactions describe positive energy states propagating forward in time.  In the basis of CP eigenstates $K_1,~ K_2$
\bea
|K_1\rangle &=& \frac{1}{\sqrt{2}}\left(|K^0\rangle + |\bar{K}^0\right) \nonumber \\
|K_2\rangle &=& \frac{1}{\sqrt{2}}\left(|K^0\rangle - |\bar{K}^0\right)
\eea
the T violation appears as an imaginary off-diagonal element
\beq
M-i\frac{\Gamma}{2} = \begin{bmatrix}
(m_1 -i\gamma_1 /2) & i{\rm Im} M_{12} \\
- i{\rm Im} M_{12} & (m_2 -i\gamma_2/2)
\end{bmatrix}
\eeq
The interference is with the $i\Gamma$ terms, such that
\beq
A(t) =\frac{2 {\rm Im}M_{12}\Delta\Gamma}{(\Delta m)^2 +(\Delta \Gamma/2)^2}
\eeq
with $\Delta m = m_S-m_L$ and $\Delta \Gamma = \Gamma_S-\Gamma_L$, where $L, S$ refer to the long-lived vs short lived states of the kaon, which in the absence of CP violation are $K_1$ and $K_2$. The point for us is that the signal comes not from the processes being really run in opposite time directions, but from the T-violating phases from the Lagrangian interfering with the phases that arise from causal propagation forward in time. Even here, Wolfenstein has pointed out that this signal is not directly T-violation, but could be faked by a T-conserving but CPT-violating interacton \cite{Wolfenstein:1999xb, Wolfenstein:1999re}.

Finally, the electric dipole moment of the neutron can be a valid test of T violation because the final state interactions are not relevant. Here the experimental measurement is again an interference with a T conserving $\exp(-iEt)$ effect, which in this case involves Larmor precession, interfering the T-conserving energy of the magnetic moment interaction with the T-violating one of the electric dipole moment interaction.

\section{Quantum Mechanics}

Now let us focus on quantum mechanics and apply canonical quantization using commutation rules. We are instructed to form the canonical momentum
\beq
p_t = \frac{\partial L}{\partial v}   = m \frac{dx}{dt}
\eeq
and from there to form the commutation relation
\beq\label{usual1}
[x, p_t] =i\hbar      ~~~~~~~~~~~~  \raisebox{.5pt}{\textcircled{\raisebox{-.9pt} {1}}} \ \ .
\eeq
When using the time variable $\tau$, we would get
the canonical momenta
\beq
p_\tau = \frac{\partial L}{\partial \tilde{v}} = m \frac{dx}{d\tau}
\eeq
which is the negative of $p_t$
\beq\label{negative}
p_t =-p_\tau.
\eeq
If one uses the same quantization assumption, we find
\beq
[x, p_\tau] =i\hbar      ~~~~~~~~~~~~  \raisebox{.5pt}{\textcircled{\raisebox{-.9pt} {2}}} \ \ .
\eeq
However, because of Eq.~(\ref{negative}), this is not the same condition as Eq.~(\ref{usual1}), but differs by a minus sign. For completeness, let us list the other possibilities
\beq
[x, p_t] = - i\hbar      ~~~~~~~~~~~~  \raisebox{.5pt}{\textcircled{\raisebox{-.9pt} {3}}} \ \ ,
\eeq
and
\beq
[x, p_\tau] = - i\hbar      ~~~~~~~~~~~~  \raisebox{.5pt}{\textcircled{\raisebox{-.9pt} {4}}} \ \ .
\eeq
If we consider the relation between $p_t$ and $p_\tau$, we see that condition \raisebox{.5pt}{\textcircled{\raisebox{-.9pt} {4}}} is equivalent to condition \raisebox{.5pt}{\textcircled{\raisebox{-.9pt} {1}}} and condition \raisebox{.5pt}{\textcircled{\raisebox{-.9pt} {3}}} is equivalent to condition \raisebox{.5pt}{\textcircled{\raisebox{-.9pt} {2}}}. These differ only in a trivial way in a relabeling of the coordinate, in the same sense as our ``joke'' in section 1 about running the clock backwards. But condition \raisebox{.5pt}{\textcircled{\raisebox{-.9pt} {1}}} and condition \raisebox{.5pt}{\textcircled{\raisebox{-.9pt} {2}}} (or \raisebox{.5pt}{\textcircled{\raisebox{-.9pt} {3}}})  are not equivalent. There remain two inequivalent possible choices, both of which are compatible with what we know about quantization.

The two choices differ in the naming of the coordinate, $t$ vs $\tau$. We know that the usual rule \raisebox{.5pt}{\textcircled{\raisebox{-.9pt} {1}}} leads to the usual description, then the alternate rules \raisebox{.5pt}{\textcircled{\raisebox{-.9pt} {2}}} or \raisebox{.5pt}{\textcircled{\raisebox{-.9pt} {3}}} will lead to the time reversed description. We can see how this arises by considering how the Hamiltonian drives time evolution in the two cases. In the usual way, the commutation rules imply that
\beq
[x, H] = \frac1{2m} ~2p_t~i\hbar = i\hbar \frac{dx}{dt} \ \ .
\eeq
This is the origin of the identification
\beq
H=i\hbar \frac{d}{dt}
\eeq
in quantum mechanics, and the fact that wavefunctions with $e^{-iEt/\hbar}$ represent positive energy solutions. With condition \raisebox{.5pt}{\textcircled{\raisebox{-.9pt} {2}}} we get instead the negative of this
\beq
[x, H] = \frac1{2m} ~2p_\tau~i\hbar = i\hbar \frac{dx}{d\tau} = -i\hbar\frac{dx}{dt} \ \ .
\eeq
In this case, Hamiltonian evolution uses
\beq
H=-i\hbar \frac{d}{dt}
\eeq
and $e^{+iEt/\hbar}$ represent positive energy solutions.

This discussion is consistent with the formal theory of time reversal. Most quantum textbooks discuss the fact that the time-reversal operation is {\em anti-unitary}. That is, it also involves a complex conjugation. As a useful example, the non-relativistic coordinate space propagator
\beq\label{NRprop}
G(\mathbf{x},\mathbf{x'}, t- t') = \sum_n \psi_n^*(\mathbf{x'})\psi_n(\mathbf{x})\exp\left[-\frac{i}{\hbar} E_n(t-t')\right]
\eeq
satisfies the time reversal property
\beq
G(\mathbf{x},\mathbf{x'}, t'- t)=G^*(\mathbf{x},\mathbf{x'}, t- t')
\eeq
This confirms that the two choices above, \raisebox{.5pt}{\textcircled{\raisebox{-.9pt} {1}}} vs \raisebox{.5pt}{\textcircled{\raisebox{-.9pt} {2}}} (or equivalently \raisebox{.5pt}{\textcircled{\raisebox{-.9pt} {1}}} vs \raisebox{.5pt}{\textcircled{\raisebox{-.9pt} {3}}}) correspond to propagation in the opposite time directions. However, at this stage we have not yet shown how these choices influence the causal structure of the theory.

Now we wish to show how the aforementioned conditions are related, that is $\raisebox{.5pt}{\textcircled{\raisebox{-.9pt} {1}}} =\fbox{1}$ and $\raisebox{.5pt}{\textcircled{\raisebox{-.9pt} {2}}} =\fbox{2}$. Let us first consider condition $\raisebox{.5pt}{\textcircled{\raisebox{-.9pt} {1}}}$. In this presentation, we follow closely the treatment by Merzbacher \cite{Merzbacher} in order that the reader may have a standard reference, although many other pathways are possible.

Suppose that a quantum system, perturbed by a constant potential $V$, is initially in an unperturbed energy eigenstate $i$ with energy $E_{i}$. If this initial state is embedded in a continuum of final states $f$, the time evolution of transition amplitudes can be derived from standard time-dependent perturbation theory:
\beq
i \hbar \frac{d}{dt} \langle f | U(t,0) | i \rangle
= \sum_{n} e^{i \frac{(E_f-E_n)}{\hbar} t} \langle f |V| n\rangle \langle n |U(t,0)| i \rangle
\label{EOM_if}
\eeq
where $E_{f} > E_{i}$, $U$ is the time-evolution operator in the interaction picture and we used that
$$
\hat{V}(t) = e^{\frac{i}{\hbar} H_{0} t} V e^{-\frac{i}{\hbar} H_{0} t}
$$
with $V$ ($\hat{V}$) being the perturbation operator in the Schr\"odinger (interaction) picture and $H_0$ is the unperturbed (time-independent) contribution to the Hamiltonian. It is clear from Eq.~(\ref{EOM_if}) that, because of the presence of several different energy gaps $(E_f-E_n)$, contributions from transition amplitudes $\langle n |U(t,0)| i \rangle$ to the equations of motion for $\langle i | U(t,0) | i \rangle$ are all of different phases, which implies that, for a continuum of final states, all such contributions tend to cancel each other. As a consequence of this destructive interference, the decay of the initial discrete state $i$ is irreversible and one cannot expect a corresponding regeneration. This reflects the presence of the arrow of causality in such quantum processes.

The equations of motion for $\langle i | U(t,0) | i \rangle$ reads
\bea
i \hbar \frac{d}{dt} \langle i | U(t,0) | i \rangle
&=& \sum_{f \neq i} e^{ -i \frac{(E_f-E_i)}{\hbar} t} \langle i |V| f\rangle \langle f |U(t,0)| i \rangle
\nn\\
&+& \langle i |V| i \rangle \langle i |U(t,0)| i \rangle .
\label{EOM_ii}
\eea
In order to derive the exponential decay law, we assume that $V$ is constant and also transitions from a discrete initial state $i$ to a quasi-continuum of final states $f$. In addition, we neglect all other contributions to the change in $\langle f | U(t,0) | i \rangle$. For $t>0$ one obtains
\beq
i \hbar \frac{d}{dt} \langle f | U(t,0) | i \rangle
=e^{ i \frac{(E_f-E_i)}{\hbar} t} \langle f |V| i \rangle \langle i |U(t,0)| i \rangle, \,\,\, f \neq i
\eeq
which has the integral form
\beq
\langle f | U(t,0) | i \rangle = - \frac{i}{\hbar} \langle f |V| i \rangle
\int_{0}^{t} dt'e^{ i \frac{(E_f-E_i)}{\hbar} t'} \langle i |U(t',0)| i \rangle .
\label{fi}
\eeq
If we substitute Eq.~(\ref{fi}) in the equations of motion for $\langle i | U(t,0) | i \rangle$ we obtain the following amplitude rate
\begin{widetext}
\beq
\frac{d}{dt} \langle i | U(t,0) | i \rangle
= \left( - \frac{1}{\hbar^2} \sum_{f \neq i} |\langle f |V| i \rangle|^2
\int_{0}^{t} dt' \, e^{ i \frac{(E_f-E_i)}{\hbar} (t'-t)}
- \frac{i}{\hbar} \langle i |V| i \rangle \right)
\langle i |U(t,0)| i \rangle
\eeq
\end{widetext}
where we removed the slowly-varying amplitude $\langle i |U(t,0)| i \rangle$ from the integrand since one is usually interested in times $t$ for which the phase factor produces rapid oscillations.

In order to do the time integration with the imaginary argument in the exponent, we need to make the integral better defined for $t\to \infty$ by using a factor of $i\epsilon$, that is
\bea\label{principal}
\int_{0}^{t} dt'  ~e^{\frac{-i}{\hbar}(E-i\epsilon)t'} &=& -i\hbar \frac{1}{E-i\epsilon}   \nonumber \\
&=& -i\hbar \left[P \frac{1}{E}+i\pi \delta(E)\right]
\eea
where $P$ denotes the Cauchy principal value. This sign is connected with the choice
$\fbox{1}$, i.e., positive energies being propagated forward in time.

The solution of the resulting differential equation is given by
\bea
\langle i | U(t,0) | i \rangle
&=& \exp \left[ -\frac{i t}{\hbar} \sum_{f \neq i} |\langle f |V| i \rangle|^2 \frac{1}{E_i-E_f + i \epsilon}
\right.
\nn\\
&-& \left. \frac{i t}{\hbar} \langle i |V| i \rangle \right] .
\label{sol_1}
\eea
In this way the use of \ref{principal} leads to
\beq
\langle i | U(t,0) | i \rangle
= \exp{ \left[ - \frac{\Gamma}{2} t - \frac{i \Delta E_{i}}{\hbar} t
 \right] }
\eeq
where
\beq
\Gamma = \frac{2\pi}{\hbar} \sum_{f \neq i} |\langle f |V| i \rangle|^2 \delta(E_f-E_i)
\eeq
and
\beq
\Delta E_{i} = \langle i |V| i \rangle
+ P\sum_{f \neq i} |\langle f |V| i \rangle|^2  \frac{1}{E_{i} - E_{f}} .
\eeq
We see that the $i\epsilon$ prescription entering here is also connected to the $i$ in the quantization condition, through the connection of positive energy propagating forward in time.  Hence this establishes the identification
$\raisebox{.5pt}{\textcircled{\raisebox{-.9pt} {1}}} =\fbox{1}$.

{This result is essentially what one would obtain if we substituted $E_n\to E_n-i\epsilon$ in the non-relativistic propagator of Eq.~\ref{NRprop}. That $i\epsilon$ and the one in the relativistic QFT propagator are clearly related.}

Now, suppose that we choose the convention associated with $\raisebox{.5pt}{\textcircled{\raisebox{-.9pt} {2}}}$. Following a similar reasoning as above, instead of Eq.~(\ref{sol_1}) we would have obtained
\bea
\langle i | U(\tau,0) | i \rangle
&=& \exp \left[ - \frac{i \tau}{\hbar} \sum_{f \neq i} |\langle f |V| i \rangle|^2 \frac{1}{E_i -E_f + i \epsilon'}
\right.
\nn\\
&-& \left. \frac{i \tau}{\hbar} \langle i |V| i \rangle \right]  .
\label{sol_2}
\eea
Now since $\tau = -t$, one must have $\epsilon' = - \epsilon$. Hence in this case one introduces a factor
$-i \epsilon$ which is connected to condition $\fbox{2}$, i.e., positive energies being propagated backward in time. As a result we confirm that $\raisebox{.5pt}{\textcircled{\raisebox{-.9pt} {2}}} =\fbox{2}$.

There is also a well known example involving causality in emission and absorption processes which highlights how the different conditions just discussed have an impact on the causal structure. This is related to the Fermi model for propagation of light in quantum electrodynamics~\cite{Fermi:32}. The Fermi's two-atom system has been discussed in detail in the literature~\cite{hamilton,fierz,ferretti,milonni1,biswas,hegerfeldt,buchholz,milonni2,power,kempf}.
The Fermi problem comprises the study of causality by means of a thorough analysis of the energy transfer between a pair of atoms. More specifically, Fermi proposes the following experiment. Consider two two-level atoms separated by some spatial distance $r$. They are coupled with a common quantum field  prepared in the Minkowski vacuum state. Suppose that, at an initial time $\tau_{0}$, atom $1$ is in the excited state and atom $2$ is in the ground state. Atom $1$ subsequently decays by emitting a photon which may in turn be absorbed by atom $2$. As a result the probability for atom $2$ to be excited remains zero until a time $\tau$ is reached such that $\tau-\tau_{0} \geq r$. For a recent discussion in the framework of disordered systems see Ref.~\cite{Menezes:17}.

The original formulation of the Fermi problem can be given using the standard techniques from time-dependent perturbation theory (within the interaction picture). For simplicity, let us consider two identical two-level atoms at rest and interacting with a massless scalar field. The atom-field interaction Hamiltonian has two local contributions of the form $\lambda\,m_{k}(\tau)\varphi[x_{k}(\tau)]$, where $\lambda$ is a small coupling constant and $m_{k}$ is the monopole moment operator of the $k$ atom. The time evolution of the system is described with respect to the proper time $\tau$ of the atoms. Assume that, at the initial time
$\tau_{0}$, the system is in the state $|\phi_{i}\rangle = |e_1 g_2\rangle \otimes |0_M\rangle$, where
$|0_M\rangle$ is the Minkowski vacuum state of the scalar field. The transition probability amplitude to the final atom-field state $|\phi_{f}\rangle = |g_1 e_2\rangle\otimes|0_M\rangle$ is given by (up to second order in $\lambda$)
\beq
{\cal A}_{\phi_{i} \to \phi_{f}}
= - \frac{\lambda^2}{4} \int_{\tau_0}^{\tau}d\tau' \int_{\tau_0}^{\tau}d\tau''
\,e^{-i\omega_0(\tau'-\tau'')}\,D_{+F}(\tau'-\tau'', r)
\label{amplitude}
\eeq
where $r = |{\bf x}_1 - {\bf x}_2|$ is the spatial separation between the atoms and $\omega_0$ is the atomic energy gap. Notice the presence of the usual Feynman propagator $D_{+F}$. Introducing the variables $\xi = \tau' - \tau''$ and $\eta = \tau + \tau'$, expression~(\ref{amplitude}) becomes
\begin{equation}
{\cal A}_{\phi_{i} \to \phi_{f}} = - \frac{\lambda^2}{4}
\int_{-\Delta\tau}^{\Delta\tau}d\xi \,(\Delta\tau - |\xi|)\, e^{-i\omega_0 \xi}\,
D_{+F}(\xi, r).
\label{amplitude2}
\end{equation}
When one inserts the expression for the Feynman propagator in Eq.~(\ref{amplitude2}), one gets two contributions. One of them will produce a causal term, proportional to the Heaviside theta function $\theta(-r + \Delta\tau)$. However, the other contribution yields a finite term for
$\Delta\tau < r$. The latter has the generic form $f(\omega_0 r)/(\omega_0^2 r^2)$, where
$f(x)$ remains bounded as $x \to \infty$. As argued by Pauli, Stueckelberg and others~\cite{Stueckelberg:50,Pauli}, the concept of causality in relativistic quantum field theory has meaning only in the wave zone
$\omega_0 r \gg 1$. Hence causality is preserved in the Fermi system.

This is the standard result. Observe that it tells us that atom $2$ will only get excited after the decay of atom $1$ (and the subsequent emission of a photon), and never before. This is enforced by the presence of the usual Feynman propagator $D_{+F}$ in the amplitude which appears as a consequence of the usual time-evolution operator (connected with the conventional causal behavior in the direction of an increasing time coordinate). However, one could also consider the time-reversed process, i.e., atom $2$ in the excited state and atom $1$ in the ground state. The associated amplitude would be the complex conjugate of Eq.~(\ref{amplitude}) since they would be related by time reversal. In this case one would employ the propagator $D_{-F}$ in the calculation. In any case, the conclusion would be the same: The ground-state atom would only get excited after the decay of the excited atom, and causality remains preserved. This simple example again clearly demonstrates the presence of an intrinsic arrow of causality in the quantum description of causal processes.

\section{Dueling arrows of causality}

It is possible for a theory to contain modes with both signs of $i$, i.e. $\fbox{1}$ and $\fbox{2}$, as long as one type of modes is unstable \cite{Donoghue:2019ecz, Lee:1969fy, Coleman, Grinstein:2008bg}. This exception illuminates the usual causality rules, by understanding the conditions under which they can be circumvented. The choice of clock direction here would be given by the stable modes. Different types of modes will have different causal directions, which means propagating against that clock direction.

A simple example of this can be seen in a theory with higher derivatives. Consider the Lagrangian for a scalar field $\phi$ with higher derivatives coupled to a normal light field $\chi$,
\bea
{\cal L} &=& \frac12 \partial_\mu\phi \partial^\mu \phi - \frac{1}{2M^2}\Box\phi \Box\phi - \kappa \chi^2 \Box\phi \nonumber  \\
&+& \frac12 \partial_\mu\chi \partial^\mu \chi \ \ ,
\eea
where $\kappa$ is a coupling constant. This can be understood more clearly using some field redefinitions. First we can use an auxiliary field $\eta$ to undo the higher derivatives, replacing this same Lagrangian with
\bea
{\cal L} &=& \frac12 \partial_\mu\phi \partial^\mu \phi - \eta \Box\phi +\frac{M^2}{2}\eta^2 - \kappa \chi^2 \Box\phi \nonumber  \\
&+& \frac12 \partial_\mu\chi \partial^\mu \chi \ \ .
\eea
This can now be separated with a shift $\phi = h-\eta$, resulting in
\bea
{\cal L} &=& \left[\frac12 \partial_\mu h \partial^\mu h  - \kappa \chi^2 \Box h\right] \nonumber  \\
&-&\left[\frac12 \left(\partial_\mu\eta \partial^\mu \eta - M^2\eta^2\right) - \kappa \chi^2 \Box\eta\right] \nonumber  \\
&+& \frac12 \partial_\mu\chi \partial^\mu \chi \ \ .
\eea
Notice the overall minus sign in the second line. When used in a path integral treatment, the first and last terms result in a normal factor of $\exp(i (S_h+S_\chi) $, but the second term enters with the opposite sign, $\exp (-iS_\eta)$. It then describes a field with a reversed arrow of causality, propagating positive energies backwards in time. Because it is a massive field coupled to light particles, it is unstable, and decays $\eta \to \chi\chi$. The overall causal structure is set by the stable fields but this extra field causes a violation of causality over a time scale of order of its width. In \cite{Donoghue:2019ecz} we have labeled this as a {\em Merlin mode} after the wizard in the Arthurian tales who ages backwards in time.

The same result can be seen without the field redefinitions by considering the propagator for the $\phi$ field. It has the behavior
\beq
i D_F(q) = \frac{i}{q^2 -\frac{q^4}{M^2} +\Sigma(q)}
\eeq
where $\Sigma (q)$ is the self energy function. For $q^2>0$, the self energy containis an imaginary part ${\rm Im} \Sigma(q) =\gamma(q)$, with $\gamma(q)> 0 $ required by unitarity. The propagator contains not only the pole at $q^2=0$, but also a massive pole at $q^2 =M^2$. Near this pole we have
\begin{eqnarray}  \label{merlin}
iD_F(q) &=& \frac{i}{q^2 - \frac{q^4}{M^2} + i \gamma(q) } \nonumber \\
&=& \frac{i}{\frac{q^2}{M^2}[M^2 - q^2 + i \gamma(q) (M^2/q^2)]} \nonumber  \\
 &\sim& \frac{-i}{q^2-M^2 -i \gamma_M} \ \ .
\end{eqnarray}
where $\gamma_M = \gamma(q^2=M^2)$. This pole is the Merlin mode. Just as a normal resonance is a finite width version of the normal propagator $D_{+F}$, the propagator near this pole is the finite width version of the time reversed propagator $D_{-F}$.

This model is an example of Lee-Wick theories \cite{Lee:1969fy}. The original Lee-Wick work was performed using canonical quantization, and the use of the different conventions on the factors of $i\hbar$ (refered to as indefinite metric quantization) were needed to quantize the Merlin models. These theories are known to violate causality at short time scales \cite{Coleman, Lee:1969fy, Grinstein:2008bg, Alvarez:2009af}. From this analysis it is clear why they do - they contain a mode which propagates with the opposite arrow of causality. Such theories are however unitary \cite{Donoghue:2019fcb}, which is made possible because the Merlin modes are unstable and do not exist as asymptotic states. The ultimate fate of such theories rests on whether the causality violation ends up being in conflict with experiment. For quadratic gravity\cite{Stelle:1976gc, Salvio:2018crh, Strumia, Donoghue:2018izj, Holdom}, which falls in this class, the causality violation happens at the order of the Planck scale and seems shielded from conflict with experiment. However, for our purposes in this paper the main point to take away is to note how the different arrows of causality can be manifest even within one theory.

\section{Conclusion}

The point of this paper is to document the way that the laws of quantum physics encodes an arrow of causality. This is contained in the factor of $i$ associated with the quantization procedure. The fundamental action or Hamiltonian may be time symmetric, but the quantization rules are not. This is most clear in the path integral framework, where the causal direction is  associated with the phase in $\exp[\pm i S]$.  This causal arrow is what gives a direction to scattering and decay processes.

However, there can also be a convention in the choice of the time coordinate, which is basically a decision to run the clock with time increasing or decreasing in the direction of the causal direction.  This implies that both possible causal directions have the same phenomenology. Either version is equally valid.

Classical physics can be said to be time-symmetric. Solutions running in both directions are equally valid, as the quote with which we started this article states. Quantum physics is not symmetric. Reactions run in one causal direction only. However, it is time-covariant, as the time reversed quantum world would have equivalent physics as the one which encodes our usual clock conventions.

These considerations have important implications for discussions of the arrow of time. The common statement saying that the fundamental laws of physics do not differentiate an arrow of time is not correct. At the microscopic level, reactions run in one direction but not the other. In a causal world there is no philosophic puzzle about the arrow of time. The laws of quantum physics give a causal direction. That direction governs the most elementary processes, such as scattering and decays. Those elementary processes when taken in large numbers govern the direction of the increase of entropy. The arrow of causality determines the arrow of thermodynamics and the arrow of time.

\section*{Acknowledgements} We would like to thank Carlo Rovelli, Lorenzo Sorbo, Ben Heidenreich, Eugene Golowich, Barry Holstein, Hermann Nicolai, Krzysztof Meissner, and  George Jorjadze for comments and conversations. GM acknowledges the hospitality of the Department of Physics, University of Massachusetts, where part of this work was carried out. The work of JFD has been partially supported by the US National Science Foundation under grant NSF-PHY18-20675 (JFD). The work of GM has been partially supported by  Conselho Nacional de Desenvolvimento Cient\'ifico e Tecnol\'ogico - CNPq under grant 310291/2018-6 and Funda\c{c}\~ao Carlos Chagas Filho de Amparo \`a Pesquisa do Estado do Rio de Janeiro - FAPERJ under grant E-26/202.725/2018.

 \end{document}